\documentclass[aps,twocolumn,nofootinbib,amssymb,amsmath,
showpacs,showkeys]{revtex4}
\newcommand{\bastar}{\begin{eqnarray*}}
\newcommand{\eastar}{\end{eqnarray*}}
\newskip\humongous \humongous=0pt plus 1000pt minus 1000pt

\newif\ifdtup

\relax
\newcommand{\be}{\begin{equation}}
\newcommand{\ee}{\end{equation}}
\newcommand{\ba}{\begin{array}}
\newcommand{\ea}{\end{array}}
\newcommand{\bea}{\begin{eqnarray}}
\newcommand{\eea}{\end{eqnarray}}
\newcommand{\pro}{\partial}

\newcommand{\hD}{{\hat D}}

\newcommand{\hR}{{\hat R}}

\newcommand{\difrac}{\displaystyle\frac}
\newcommand{\nn}{\nonumber}

\begin{document}
\title{Non-Topological Gauss-Bonnet type model of gravity with torsion}
\bigskip
\author{H. Niu}
\email{hniu1982@hotmail.com}
\affiliation{S. S. Chern Institute
of Mathematics, Nankai University, Tianjin, 300-071, China}
\author{D. G. Pak}
\email{dmipak@phya.snu.ac.kr}
\affiliation{S. S. Chern Institute
of Mathematics, Nankai University, Tianjin, 300-071, China}
\affiliation{ Center for Theoretical Physics, Seoul National
University, Seoul 151-742, Korea}
\begin{abstract}
A non-topological Lorentz gauge model of gravity
with torsion based on Gauss-Bonnet type Lagrangian is considered.
The Lagrangian differs from the Lovelock term in four-dimensional
space-time and has a number of interesting features.
We demonstrate that the model
admits a propagating torsion unlike the case
of the topological Lovelock gravity.
Due to additional symmetries of the proposed Gauss-Bonnet type
Lagrangian the torsion has a reduced set of dynamical
degrees of freedom corresponding to the spin two field,
$U(1)$ gauge vector field and spin zero field.
A remarkable feature is that the kinetic part of
the Hamiltonian containing the spin two field
is positively defined. We perform one-loop
quantization of the model
for a special case of constant Riemann curvature space-time
background treating the torsion as a quantum field variable.
We discuss a possible mechanism of emergent Einstein
gravity as an effective theory which can be
induced due to quantum dynamics of torsion.
\end{abstract}

\maketitle

\section{Introduction}

Recently a modified Gauss-Bonnet gravity is a subject of
intensive studies in constructing alternative cosmological
models \cite{GBcosm1, GBcosm2}. The Gauss-Bonnet term appears also
in low-energy effective action of superstring \cite{string}
which is a candidate for a consistent theory of quantum gravity
unified with other fundamental interactions.
On the other hand, the gauge approach to gravity based on
gauging Lorentz and Poincare groups \cite{uti, kibble,transl,ivan2,cho11}
can also lead to a consistent quantum theory
of gravity in the framework of field theory formalism
\cite{anttomb,carmeli1,martell}.
The extension of gravity models to the case
of non-Riemannian space-time geometry reveals new possibilities
towards construction of renormalizable quantum gravity with torsion
\cite{ivan1,hehl,odints,shapiro}.
A Lorentz gauge model of gravity with Yang-Mills type Lagrangian
including torsion has been developed further in \cite{pak}
where it has been proposed that the Einstein gravity
with a cosmological term can be induced as an effective theory
due to quantum corrections of torsion.
In that model the space-time metric is treated as a fixed classical
field while the contortion (torsion) supposed to be a quantum field.
Such a treatment of the metric is not satisfactory from the
conceptual point of view since one has to assume the existence
of the fixed classical space-time with a metric given a priori. In other words,
we encounter the problem of space-time background dependence which is similar
to the space-time dependence problem in superstrings.
One possible way to resolve this problem is to generalize the
Lorentz gauge model by extending the gauge group to Poincare one.
In that case the gauge potential of Poincare group, the vielbein,
becomes dynamical on equal footing with torsion.
Another interesting possibility is to consider
such a gravity model which assumes the existence
of a pure topological phase with unfixed,
arbitrary metric from the start.

    In the present paper we consider a non-topological Gauss-Bonnet
type gravity model with torsion and study its classical and quantum
properties. We demonstrate that the model admits a propagating torsion,
and this is strictly different from the Gauss-Bonnet gravity model
in Lovelock form {\cite{love,zanelli}}.
The non-topological Gauss-Bonnet gravity
represents an alternative theory to Yang-Mills
type gauge model of  gravity
and it can provide the mechanism
of emergent Einstein gravity via quantum dynamics of torsion
as it was proposed recently in \cite{pak}.
The main advantage of the present model is that
in the absence of torsion the theory describes a pure topological phase
of gravity with an arbitrary space-time metric which does not
satisfy any equation of motion. The metric becomes dynamical only
after inducing the Einstein-Hilbert term
in the quantum effective action. It is remarkable
that the torsion in our model possesses a reduced set
of physical field components including a unique spin two field.
So that the torsion can be interpreted as a gravitational quantum
counter-part to the metric treated as a classical
field variable of the effective Einstein gravity.
Besides, the spin two field component of torsion leads to
a positively-definite kinetic term in the Hamiltonian,
unlike the case of $R^2$ Lorentz gauge gravity models
which have the well-known non-unitarity problem .

In section II, we describe the non-topological model of
Gauss-Bonnet type gravity in the framework of Riemann-Cartan geometry.
In section III, we study the canonical structure of the model
in a special case of flat space-time metric and non-vanishing torsion
which plays a role of the gauge potential in Lorentz gauge field theory.
The quantization of the model in a constant curvature space-time background
with quantum torsion field is considered in section IV.
In the last section we discuss the possibility of generating
the Einstein-Hilbert term and cosmological constant as a result of
torsion radiative corrections.

\section{The Model}

Let us start with the main outlines of Riemann-Cartan geometry.
The basic geometric objects in approaches to formulation of gravity as a gauge
theory of the Poincare group \cite{uti,kibble,cho11} are
the vielbein $e_a^m$ and the general Lorentz affine connection $A_m^{~cd}$.
The infinitesimal Lorentz transformation of the
vielbein $e_a^m$ is given by
\bea
&&\delta e_a^m= [{\bf \Lambda},e_a^m]=\Lambda_a^{~b} e_b^m,
\eea
where ${\bf \Lambda} \equiv \Lambda_{cd} \Omega^{cd}$ is a Lie algebra
valued gauge parameter, and $\Omega^{cd}$ is a generator of Lorentz
Lie algebra. We assume that the vielbein is
invertible and the signature of the flat metric $\eta_{ab}$ in the tangent
space-time is Minkowskian, $\eta_{ab}={\rm diag} (+---)$.

The covariant derivative with respect to Lorentz group
transformation is defined in a standard manner
\bea
D_a=e_a^m (\pro_m + g{\bf A}_m) ,
\eea
where ${\bf A}_m\equiv A_{m cd} \Omega^{cd}$ is a general
affine connection taking values in the Lorentz Lie algebra, and
$g$ is a new gravitational gauge coupling constant.
For brevity of notation we will use a
redefined connection which absorbs the coupling constant. The
original Lorentz gauge transformation of the connection ${\bf A}_m$ has the form
\bea
 \delta {\bf A}_m&=&-\pro_m {\bf \Lambda} - [{\bf A}_m, {\bf \Lambda}].
\label{eq:deltaA}
\eea

The affine connection $A_{m cd}$ can be rewritten
as a sum of Levi-Civita spin connection $\varphi_{m c}^{~~d}(e) $
and contortion $K_{m c}^{~~d}$
\bea
A_{m c}^{~~d} &=& \varphi_{m c}^{~~d} + K_{m c}^{~~d}, \label{split}\\
\varphi_{\mu a}^{~~b}(e)&=&\difrac{1}{2} (e^{\nu b} \pro_\mu e_{\nu a}-
e_a^\nu e_\mu^c \pro^b
e_{\nu c}+\pro_a e_\mu^b \nn \\
&&-e_a^\nu \pro_\mu e_\nu^b
+e^{\nu b} e_\mu^c \pro_a e_{\nu c}-\pro^b e_{\mu a}).\label{LeviCivita}
\eea

The torsion and curvature tensors are defined in a standard way
\bea
&& [D_a,D_b]=T_{ab}^{\,\,c} D_c+{\bf R}_{ab} , \nn \\
&& T_{ab}^{\,\,c}=K_{ab}^{~c}-K_{ba}^{~c},
\eea
here, ${\bf R}_{ab} \equiv R_{ab cd} \Omega^{cd}$.
Under the decomposition (\ref{split})
the Riemann-Cartan curvature is splitted into two parts
\bea
&& R_{abcd}=\hat R_{abcd}+\tilde R_{abcd},  \\
&& \hat R_{abcd}=\hat D_{\underline a} \varphi_{\underline b \underline c}^{
~\,\underline d}+\varphi_{ac}^{~~\,\,e}\varphi_{be}^{~\,d}-(a\leftrightarrow b), \nn \\
&& \tilde R_{abcd}=\hat D_{\underline a} K_{\underline b \underline c}^{
~\,\underline d}+K_{ac}^{~~e} K_{be}^{~\,d}-(a\leftrightarrow b) , \nn
\eea
where the underlined indices stand for indices over which the
covariantization has been performed.

With these preliminaries let us write down the well-known
Lagrangian of Gauss-Bonnet gravity of Lovelock type in
four-dimensional space-time \cite{love,zanelli}
\bea
&&  {\cal L}_{Lovelock}= \beta_0 \epsilon_{abcd} R^{ab}\wedge R^{cd}, \label{Lovelock}
\eea
where $\beta_0$ is a dimensionless constant and $R_{ab}$
is the Riemann-Cartan curvature two-form.
Since the Lagrangian is represented by a closed differential
form it does not produce equations of motion,
so that the torsion has no propagating modes.
We will consider a model of gravity with torsion
based on the following Gauss-Bonnet type Lagrangian
\bea
{\cal L}=-\dfrac{1}{4} I_{GB}=
-\difrac{1}{4} (R_{ab cd} R^{abcd}-4 R_{ab} R^{ab}
 + R^2). \label{L0}
\eea
In a case of Riemannian space-time geometry, when torsion
vanishes, the Lovelock term, (\ref{Lovelock}), reduces
to the standard Gauss-Bonnet
topological invariant which can be
rewritten in its original form $I_{GB}$, (\ref{L0}),
in terms of Riemann curvature (up to a normalization factor).
However, it is important to stress that in the presence of torsion
the Gauss-Bonnet combination $I_{GB}$ is
principally different from the Lovelock term ${\cal L}_{Lovelock}$.
Namely, one can check that the Gauss-Bonnet Lagrangian
in the form (\ref{L0}) does not correspond to any topological invariant
in Riemann-Cartan geometry since it can not be expressed
as a total divergence.
In the subsequent sections we will demonstrate
that the contortion (torsion) in our model with Lagrangian
(\ref{L0}) reveals non-trivial dynamical properties and a number
of interesting features.

\section{Gauss-Bonnet type gauge model in flat space-time}

The Lorentz gauge model described by Gauss-Bonnet type Lagrangian (\ref{L0})
contains two sets of variables, the vielbein
and the torsion. As we will see in the next section,
the one-loop effective action with a constant curvature
space-time background and quantum torsion possesses additional
local symmetries which reduce the dynamical
content of torsion.
However, beyond one-loop approximation and with no
assumption of constant curvature space-time
the local symmetries may not survive. So that
it is not obvious whether the number of dynamical torsion
degrees of freedom remains the same in general.
To understand the origin of the dynamical content of torsion
we consider first the classical structure
of Gauss-Bonnet gravity in the simplest case when
the space-time metric is flat. In this limit the model
represents a pure Lorentz gauge field theory
with the quadratic Lagrangian  ${\cal L}$, (\ref{L0}),
and contortion $K_{bcd}$ as a gauge potential.
The Riemann-Cartan curvature represents
the Lorentz gauge field strength
which can be written in a standard form (we keep
for the Lorentz gauge field strength the same notation as for the
Riemann-Cartan curvature)
\bea
R_{abcd}=\pro_a K_{bcd}-\pro_b K_{acd}
+K_{ac}^{~~e}K_{bed}-K_{bc}^{~~e}K_{aed}.
\eea

Let us consider the canonical structure of the Lorentz
gauge theory along the lines of canonical formalism
of theories with constraints \cite{gitman}.
The canonical momenta corresponding to the gauge potential
$K_{mcd}$ are defined as follows
\bea
 \pi_{mcd}=\dfrac{\pro {\cal L}}{\pro \dot K^{mcd}}&=&
    -R_{0mcd}+(\delta_{0c} R_{md}+\delta_{md} R_{0c}\nn \\
    && +\dfrac{\gamma}{2} R \delta_{0d}\delta_{mc}-(c\leftrightarrow d)),
\eea
where $\dot K^{mcd}=\pro_0 K_{mcd}$.
One can check that the canonical momenta $\pi_{0cd}$
vanish identically and represent first-stage
constraints
\bea
&&\pi_{0cd}=0. \label{con11}
\eea
There are other momenta, $\pi_{\mu 0\delta}$ (we use Greek letters
for space indices), which do not contain
terms with time derivatives of $K_{mcd}$ and,
due to this, they produce additional primary constraints
\cite{gitman}
\bea
&& \pi_{\mu 0\delta}=-R_{0\mu 0\delta}+R_{\mu\delta}+\delta_{\mu\delta} R_{00}-
\dfrac{1}{2}\delta_{\mu\delta} R.
\eea
In Lagrange formalism it follows that the Lagrange equations
of motion for the field $K_{\mu 0 \delta}$
are not dynamical but represent first order differential
equations. These equations are solvable constraints in the theory
which can be solved for $K_{\mu 0 \delta}$
at least in principle. So that the components $K_{\mu 0\delta}$
do not represent dynamical degrees of freedom,
and they can be excluded from the physical field spectrum.

The remaining canonical momenta $\pi_{\mu\gamma\delta}$
have the following form
\bea
&& \pi_{\mu\gamma\delta}=-R_{0\mu\gamma\delta} +\delta_{\mu\delta} R_{0\gamma}
-\delta_{\mu\gamma} R_{0\delta}, \nn \\
&& \pi^\mu_{~\mu\delta}\equiv\pi_\delta=-R_{0\delta}.
\eea
These equations can be resolved to find the "velocities"
$\dot K_{\mu\gamma\delta}$
\bea
\dot K_{\mu\gamma\delta}&=&-\pi_{\mu\gamma\delta}+\delta_{\mu\gamma} \pi_\delta
             -\delta_{\mu\delta} \pi_\gamma \nn \\
&&+\pro_\mu K_{0\gamma\delta}-K_{0\gamma}^{~~c}K_{\mu c \delta}+K_{0\delta}^{~~c}K_{\mu c \gamma},\nn \\
\dot K^\mu_{~\mu\delta}&=&\pi_\delta+\pro^\nu K_{0\nu\delta}+K^{c} K_{0c\delta}-K_{0ce}K^{ce}_{~~\delta},
\label{resolvs1}
\eea
where $K^c=K_b^{~bc}$.
One can insert the "velocities" $\dot K_{\mu\gamma\delta}$
into the initial Lagrangian
\bea
{\cal L}&=&-\dfrac{1}{2} \pi_{\mu\gamma\delta}^2 +\pi_\delta^2 +2 \dot K^{\mu 0 \delta}
(-R_{0\mu 0\delta}+R_{\mu\delta} \nn \\
    &+& \delta_{\mu\delta} R_{00}-
\dfrac{1}{2}\delta_{\mu\delta} R)+\check {\cal L}(K),\nn \\
 \check {\cal L}(K)&=&-\dfrac{1}{4} R^2_{\mu \beta cd}-\check R^2_{0\beta 0\delta}
+\check R^2_{\beta \delta}+\check R^2_{\beta 0}\nn \\
&+&\check R^2_{00}-\dfrac{1}{4} \check R^2 ,
\eea
where $\check R_{abcd}$ are defined by the expressions for the field strength
$R_{abcd}$ with omitted time derivative terms.
With this one can define an extended Hamiltonian \cite{gitman}
\bea
H^{*}=\pi_{mcd} \dot K^{mcd}-{\cal L}.
\eea
After inserting the functions $\dot K_{\mu\gamma\delta}$,
(\ref{resolvs1}), into the  extended Hamiltonian $H^*$
one obtains the Hamiltonian $H^{(1)}$ with the partially resolved
"velocities" $\dot K_{\mu\gamma\delta}$
and first-stage constraints $\Phi^{(1)}_{1cd}, \Phi^{(1)}_{2\mu\delta}$
\bea
 H^{(1)}&=&\dfrac{1}{2} \vec \pi_{\mu\gamma\delta} \vec \pi_{\mu\gamma\delta}
-\vec \pi^\delta \vec \pi^\delta
 -\pi^{\mu\gamma\delta} \check R_{0\mu\gamma\delta}\nn \\
&&-\check {\cal L}(K)
+\lambda^{0cd}\Phi^{(1)}_{1cd} + 2 \lambda^{\mu 0 \delta}\Phi^{(1)}_{2\mu\delta},\nn\\
\Phi^{(1)}_{1cd}& =& \pi_{0cd}, \nn \\
\Phi^{(1)}_{2\mu\delta} &=&\pi_{\mu 0\delta}+\check R_{0\mu 0\delta}-
\check R_{\mu\delta}\nn \\
&&+\dfrac{1}{2} \delta_{\mu\delta} (\check R-2 \check R_{00}),
 \eea
 where $\lambda^{0cd}, \lambda^{\mu 0 \delta}$ are Lagrange
 multipliers. We can see that the first kinetic term
 in the Hamiltonian provides a positive
 contribution to the energy.
There is no term quadratic in momentum $\pi_{\mu 0\delta}$
which could produce a negative energy contribution as in the case
of Yang-Mills type $R^2$-gravity. This is a direct consequence
of the specific structure of the Gauss-Bonnet combination.
Still one has a negative contribution coming from the second term in $H^{(1)}$,
so that the total Hamiltonian is not positively-defined.
Notice, the Weyl type Lagrangian
\bea
&&  {\cal L}_{Weyl}=-\dfrac{1}{4}(
R_{ab cd} R^{abcd}-2 R_{ab} R^{ab}
 +\dfrac{1}{3} R^2) \label{Weyl}
\eea
leads to a Hamiltonian with a non-vanishing kinetic term
$-\vec \pi_{\mu 0 \delta}^2$, whereas the kinetic term
$\pi_d^2$ does not appear at all since the canonical momenta $\pi_d$
vanish identically.

 One can verify that the first-stage constraints
 commute to each other
\bea
\{\Phi^{(1)}_{1cd}, \Phi^{(1)}_{2\mu\delta}\}&=&0.
\eea
By direct calculating the Poisson brackets between the Hamiltonian
$H^{(1)}$ and the first stage constraints $\Phi^{(1)}_{1 cd}$
one can find the second-stage constraints
 \bea
\{H^{(1)},\Phi^{(1)}_{1\gamma\delta}\}= \Phi^{(2)}_{1\gamma\delta}&=&
  D^i \pi_{i\underline\gamma\underline\delta}+K^\beta_{~0\gamma} (\Phi^{(1)}_{2\beta\delta}-
\pi_{\beta 0 \delta}), \nn \\
\{H^{(1)},\Phi^{(1)}_{10\delta}\}= \Phi^{(2)}_{10\delta}
&=&\pi_{\mu\gamma\delta} K^{\mu\gamma}_{~~\,0}+\pro^\beta
  (\Phi^{(1)}_{2\beta\delta}-\pi_{\beta 0\delta}) \nn \\
  &-&K^{\beta\gamma}_{~~\,\delta}
  (\Phi^{(1)}_{2\beta\gamma}-\pi_{\beta 0\gamma}),
  \eea
  where
 the covariantization is assumed on underlined indices.
 One can calculate the following Poisson bracket
\bea
&& \{\Phi^{(2)}_{1cd}, \Phi^{(1)}_{1ab}\}=0.
\eea

This implies that the Lagrange multiplier $\lambda_{0cd}$
can not be found as a solution of a new constraint.
This is consistent with the fact of presence of the
original Lorentz gauge symmetry due to which one can
impose the Coulomb gauge condition $K_{0cd}=0$.

The explicit expression
for the second-stage constraint $\Phi^{(2)}_{2\mu\delta} $
turns out to be quite complicate.
Because of this it is hard to find an explicit solution
for the second Lagrange multiplier $\lambda_{\mu 0\delta}$
as a solution of higher stage constraints.
This obstacle reflects the high non-linearity
of the Lagrange equation for $ K_{\mu 0 \delta}$.
Notice that, since $K_{\mu 0 \delta}$ satisfies
first order differential equation, it
can not be set to zero. That means there is no additional
local symmetries in the full Lagrangian except for the original
one given by (\ref{eq:deltaA}).
To analyze the number of local dynamical degrees
of freedom in the theory
it is enough to consider a free part of the Lagrangian (\ref{L0})
\bea
{\cal L}_{free}&=&
-\dfrac{1}{2} (\pro_a K_{bcd})^2 +\dfrac{1}{2} (\pro^b K_{bcd})^2
+(\pro^a K_{bad})^2 \nn \\
&-&2 \pro^a K_{bad} \pro^b K^d +(\pro_b K_d)^2-(\pro^b K_b)^2.
\eea
One can verify that there are
indeed only nine physical degrees of freedom corresponding
to the field $K_{\mu\gamma\delta}$.
To see that, notice that the free Lagrangian
has additional two types of gauge symmetries
under the following transformations
\bea
&& \delta K_{mcd}=\pro_c \tau_{dm}-\pro_d \tau_{cm},  \label{symtau}\\
&&\delta K_{mcd}=\pro_c \sigma_{dm}-\pro_d \sigma_{cm}, \label{symsigma}
\eea
where $\tau_{cd}, \sigma_{cd}$
are constrained gauge parameters satisfying
the conditions
\bea
&&\sigma_{cd}=\sigma_{dc}, ~~~~\pro^c \sigma_{cd}=0, \nn \\
&&\tau_{cd}=-\tau_{dc}, ~~~~\pro^c \tau_{cd}=0.
\eea
The constrained gauge parameter $\sigma_{cd}$ has six independent
degrees of freedom whereas the parameter $\tau_{cd}$ has
only three independent degrees of freedom. To count the
independent degrees of $\tau_{cd}$ it is convenient to
replace $\tau_{cd}$ with its dual counter-part
\bea
&& \tau_{cd} = -\dfrac{1}{2} \epsilon_{cdab} \tau^{*ab}.
\eea
This replacement allows to express the dual gauge
parameter $\tau^*_{cd}$ in terms of a new vector $\psi_a$
\bea
&& \tau^*_{ab}=\hD_a \psi_b-\hD_b \psi_a. \label{param}
\eea
The definition of $\psi_a$ implies a secondary gauge invariance
\bea
\delta_\chi \psi_a=\hD_a \chi \label{secpar}
\eea
which decreases the number of independent degrees of freedom
up to three.
So that, the total number of independent pure gauge degrees of freedom
for the symmetries (\ref{symtau}, \ref{symsigma}) is nine.
After subtracting six Lorentz gauge degrees
we obtain finally nine physical degrees
of freedom for $K_{bcd}$ in agreement with
the results obtained in the canonical formalism.

In the next section we will  consider the quantization
of the Gauss-Bonnet gravity model in the presence of non-flat
background metric corresponding to a constant curvature space-time.
We will show that the number of additional local gauge
symmetries will be decreased, however, the number of
physical dynamical degrees of freedom remains the same.

\section{Quantization in a constant curvature space-time background}

One-loop effective action with a constant curvature space-time
background and quantum torsion has been calculated recently
in the model with Yang-Mills type Lagrangian quadratic
in Riemann-Cartan curvature \cite{pak}.
In this section we consider perturbative quantization of the model
with Gauss-Bonnet type Lagrangian, (\ref{L0}). The quantization procedure
is similar to the covariant background quantization
in supergravity \cite{pak2}. We apply the quantization scheme based on
functional integral.
In background field formalism one starts with splitting
the general gauge connection $A_{m cd}$
into background (classical) and quantum parts
\bea
&A_{m cd}= A^{({\it cl})}_{m cd}+A^{({\it q})}_{m cd}. \label{splitting}
\eea
In this section
we identify the classical field $A^{{\it (cl)}}_{m cd}$
with the Levi-Civita connection $\varphi_{mcd}(e)$
corresponding to the Riemannian space-time geometry
and the quantum part $A^{{\it (q)}}_{m cd}$
with contortion $K_{m cd}$ which represents
quantum dynamical degrees of freedom.

Let us define two types of Lorentz
gauge transformations consistent with the
original gauge transformation (\ref{eq:deltaA}) and splitting
(\ref{splitting}): \newline
(I) the classical, or background, gauge transformation
\bea
&&\delta e_a^m = \Lambda_a^{~b} e_b^m, \nn \\
&& \delta {\boldsymbol \varphi}_m(e) = -\pro_m {\bf \Lambda}-
                 [{\boldsymbol \varphi}_m,{\bf \Lambda}], \nn\\
&& \delta {\bf K}_m = -[{\bf K}_m,{\bf \Lambda}],\label{eqI}
\eea
(II) the quantum gauge transformation
\bea
&&\delta e_a^m =
\delta {\boldsymbol \varphi}_m(e)=0, \nn \\
&& \delta {\bf K}_m= - \hat D_m {\bf \Lambda}-[{\bf K}_m,{\bf \Lambda}],\label{eqII}
\eea
where ${\boldsymbol \varphi}_m \equiv \varphi_{m cd} \Omega^{cd}$, and
the restricted covariant derivative $\hat D_m$ is defined
by means of the Levi-Civita connection only
\bea
&&\hat D_m {\bf\Lambda}=\pro_m {\bf\Lambda}+[{\boldsymbol \varphi}_m,{\bf \Lambda}].
\eea
Notice that the restricted derivative $\hat D_m$ is covariant under
the classical Lorentz gauge transformation.

In one-loop approximation it is sufficient to keep
only quadratic contortion terms in the Lagrangian.
After integration by part and neglecting surface terms
the quadratic Lagrangian can be reduced to the form
\bea
&&{\cal L}^{(2)}=-\dfrac{1}{4}[I_{GB}(\hR)+2\hD_a K_{bcd}(\hD^a K^{bcd}-\hD^b K^{acd})\nn\\
&&-4 (\hD^a K_{bad})^2 + 8 \hD^a K_{bad} \hD^b K^d -4 (\hD_b K_d)^2 \nn \\
&&+4 (\hD_b K^b)^2 + 4 \hat R_{abcd} K^{ace}K^{b~d}_{~e}
-8 \hat R^{bd} (K^e K_{bed}\nn\\
&&-K_{bce}K^{ce}_{~~d})-2 \hat R (K_d^2+K^{bce} K_{ceb})], \label{lagrini}
\eea
where $I_{GB}(\hR)$ is the topological
Gauss-Bonnet density (up to an appropriate normalization factor).

To simplify the analysis of local dynamical degrees of freedom in the theory
with curved space-time background (within the framework of perturbative quantization)
we will use adiabatic approximation. So that the Riemann curvature
is supposed to be covariant constant, i.e.,
$\hat D_a \hR_{bcde}=0$.

An interesting feature of the quadratic Lagrangian (\ref{lagrini})
is the presence of additional local $U(1)$ symmetry
\bea
&& \delta_{U(1)} K_{bcd} = \dfrac{1}{3} (\eta_{bc}\hD_d\lambda-\eta_{bd}\hD_c \lambda),\nn \\
&& \delta_{U(1)} K_{d} = \hD_d \lambda.  \label{U1}
\eea
This $U(1)$ symmetry corresponds to the symmetry (\ref{symsigma}) in the flat
space-time with the gauge parameter $\lambda\equiv\sigma^{~c}_c$.
Notice, that a free part of the Yang-Mills type Lagrangian
\bea
&& {\cal L}_{YM}=-\dfrac{1}{4} R_{abcd}^2 \label{LYM}
\eea
does not possess such a local $U(1)$ symmetry.

It is convenient to decompose the contortion into
irreducible parts
\bea
&& K_{bcd}=Q_{bcd}+\dfrac{1}{3} (\eta_{bc} K_d-\eta_{bd} K_c)
+ \dfrac{1}{6} \epsilon_{bcde} S^e, \nn \\
&& Q^c_{~cd}=0,  \nn \\
&& \epsilon^{abcd} Q_{bcd}=0.
\eea
For simplicity we choose the covariant constant background space-time
as a Riemannian space-time of constant curvature
\bea
\hat R_{abcd}= \difrac{1}{12} \hat R (\eta_{ac} \eta_{bd}-\eta_{ad}\eta_{bc}).
\label{background}
\eea
With this the Lagrangian (\ref{lagrini}) can be rewritten in the form
\bea
&&{\cal L}^{(2)} =-\dfrac{1}{4} I_{GB}(\hR)
-\dfrac{1}{2} (\hD_a Q_{bcd})^2
 +\dfrac{1}{2} \hD^a Q^{bcd} \hD_b Q_{acd} \nn \\
 &&+(\hD^a Q_{bad})^2-\dfrac{2}{3}\hD^a Q_{bad} \hD^b K^d
 +\dfrac{1}{12} \hat R Q^{bce}Q_{ceb} \nn \\
 &&+\dfrac{1}{9} (\hD_b K_d-\hD_d K_b)^2
 +\dfrac{1}{12} (\hD_a S^a)^2 -\dfrac{1}{36} \hR S^2. \label{Lquadr}
\eea

An unexpected feature of the quadratic Lagrangian ${\cal L}^{(2)}$,
(\ref{Lquadr}), is that it admits another local symmetry
corresponding to the symmetry (\ref{symsigma})
in flat space-time limit.
The symmetry is provided by the following
transformations with a new constrained parameter
$\chi_{bc}$
\bea
&& \delta_\chi Q_{bcd}= \hD_c \chi_{db}-\hD_d \chi_{cb}, \nn \\
&& \delta_\chi K_d=0, \nn \\
&& \delta_\chi S^a=0, \nn \\
&& \chi_{bc}=\chi_{cb},~~~\chi^c_{~c}=0, ~~~\hD_c\chi_{cd}=0. \label{eqchi}
\eea
The field $Q_{bcd}$ has sixteen field components in general.
After subtracting six pure gauge degrees
of freedom due to Lorentz gauge symmetry and five degrees
due to $\chi$-symmetry one has exactly five physical
degrees of freedom for the spin two field.
The fact that we have only one physical spin two field
is unexpected, and it does not occur in the gauge gravity model with Yang-Mills type
Lagrangian ${\cal L}_{YM}$, (\ref{LYM}), where the contortion
$Q_{bcd}$ contains a pair of two spin fields, one of which
produces a negative contribution to the Hamiltonian.

Notice, that the last term in the Lagrangian (\ref{Lquadr})
prevents appearance of the local symmetry (\ref{symtau}).
This is not surprising, because
the symmetries available in the case of flat space-time
may not survive in curved space-time theory.
Nevertheless, one can easily verify from the equations
of motion for the vector field $S^a$ that only the temporal
component $S^0$ is dynamical.
The equations of motion for the space field components $S^\mu$
represent the first-order differential equations in respect to time derivative,
so that they are constraints in the theory which can be solved.
This implies that the field $S_a$ contains only one dynamical
degree of freedom corresponding to spin zero field.

The fact that the field $S_a$ has only scalar dynamical
degree of freedom completes the analogy between
the metric tensor which has spin 2 and spin 0 irreducible
components and the torsion fields $Q_{bcd}, S_a$ which
have exactly six dynamical degrees of freedom corresponding to
fields of spin two and zero.
This can serve as an additional argument to our
conjecture that the torsion represents a dynamic degree of
freedom of quantum gravity \cite{pak} and the classical metric tensor
inherits its properties after the Einstein gravity emerges
as an effective theory via quantum dynamics of torsion.
We will discuss on this in more details in the last section.
Notice, we have totally nine physical degrees of freedom
for the fields $Q_{bcd}, K_d, S^a$ in agreement with the
analysis presented in the previous section.

Having all local symmetries of the quadratic Lagrangian
one can perform the formal quantization using the standard
methods of quantum field theory.
First we fix the gauge under the quantum type (II) gauge
transformations, (\ref{eqII}), which can be written
for the torsion irreducible fields $Q_{bcd}, K_a, S_a$
\bea
\delta Q_{bcd}&=& \delta K_{bcd} -\dfrac{1}{3}(\eta_{bc} \delta K_d
     -\eta_{bd} \delta K_c)
      -\dfrac{1}{6} \epsilon_{bcde} \delta S^e, \nn \\
 \delta K_d&=&-\hD^c \Lambda_{cd}, \nn \\
 \delta S^a &=&-\epsilon^{abcd} \hD_b \Lambda_{cd},
\eea
where we keep only linear terms, that is
enough in one-loop approximation.
The simplest gauge fixing function
we have chosen is the following
\bea
&&F_{1cd}=\hD^b Q_{bcd}.
\eea
One has a simple transformation rule for the gauge function
\bea
&& \delta F_{1cd} = -\dfrac{2}{3} ( \hD\hD+\dfrac{\hR}{6}) \Lambda_{cd}.
\eea
With this one can write down
the corresponding gauge fixing term and Faddeev-Popov ghost
Lagrangian
\bea
&& {\cal L}^{(1)}_{gf}=-\dfrac{1}{2 \xi_1} (\hD^b Q_{bcd})^2,  \nn \\
&& {\cal L}^{(1)}_{FP}=\bar c_1^{cd} (\hD\hD+\dfrac{\hR}{6}) c_{1cd},
\eea
where $\bar c, c$ are ghost fields.
For simplicity we choose the gauge parameter $\xi_1=1$.
Notice, that the gauge function $F_{1cd}$ is invariant under
$U(1)$ and $\chi$-transformations, (\ref{U1}, \ref{eqchi}).
To fix the gauge for the local $U(1)$ symmetry
one has to introduce a second gauge fixing function
which can be chosen as
\bea
F_2=\hD^b K_b.
\eea
The corresponding gauge fixing term and Faddeev-Popov Lagrangian
have the following form
\bea
&& {\cal L}^{(2)}_{gf} = -\dfrac{1}{\xi_2} (\hD^b K_b)^2, \nn \\
&& {\cal L}^{(2)}_{FP} = \bar c_2 \hD\hD c_2.
\eea
We choose a Feynman gauge ($\xi_2=1$) for simplicity.
Finally, to fix the gauge for $\chi$-transformations one
can choose the following gauge fixing function
\bea
&& F_{3bd}=\dfrac{1}{2}(\hD^a Q_{bad}+\hD^a Q_{dab}), \nn \\
&& \eta^{bd}F_{3bd}=0, \nn \\
&& \hD^b F_{3bd}=\dfrac{1}{2} \hD^a\hD^b Q_{bad} \simeq 0,
\eea
where the last equality takes place on the hypersurface $\hD^b Q_{bcd}=0$
in the configuration space of functions $\{Q_{bcd}\}$.
One can easily find the corresponding gauge fixing and ghost terms
\bea
&& {\cal L}^{(3)}_{gf}=-\dfrac{1}{2\xi_3} F_{3bd}^2, \nn \\
&& {\cal L}^{(3)}_{FP}=\bar \psi^{cd} (\hD\hD-\dfrac{\hR}{3}) \psi_{cd}. \label{tachyon}
\eea
We set $\xi_3=\dfrac{1}{2}$ which corresponds to
symmetric gauge.

The final expression for a total one-loop
effective Lagrangian is given by the sum
of all gauge fixing and ghost terms
\begin{widetext}
\bea
&& {\cal L}^{(2)}_{eff} =-\dfrac{1}{4} I_{GB}(\hR)-
\dfrac{1}{2} (\hD_a Q_{bcd})^2 +\dfrac{1}{2}(\hD^a Q_{bad})^2
-\dfrac{1}{2}\hD^a Q_{bad} \hD_c Q^{dcb}
 -\dfrac{2}{3} \hD^a Q_{bad} \hD^bK^d \nn \\
&& -\dfrac{1}{8} \hR Q_{bcd}^2
  +\dfrac{\hR}{6}Q^{bcd}Q_{cdb}+
\dfrac{2}{9} (\hD_a K_b)^2 + \dfrac{1}{18} \hR K_d^2
-\dfrac{1}{36} S^a (3 \hD_a \hD_b+\eta_{ab} \hR) S^b
 +\sum_{i=1,2,3}{\cal L}^{(i)}_{FP}.\label{eq:L2}
\eea
\end{widetext}
The expression for the effective Lagrangian is ready for
calculation of the one-loop effective action. The contributions
of ghosts  are given by scalar functional
determinants which can be easily calculated in analytic form as in \cite{pak}.
Calculation of the contribution produced by contortion is much more complicate due to
the tensorial structure of the corresponding propagator. At least it can be estimated
by using the high derivative expansion.

The propagator for the fields $Q_{bcd}, K_d$ can be
found straightforward. The principal part of such calculation
is to find an inverse operator to operator
${\cal G}_{bcd}^{pqr}$ in the kinetic term
\bea
&& {\cal L}^{\it kin} (Q)= \dfrac{1}{2} Q^{bcd}{\cal G}_{bcd}^{pqr} Q_{pqr}.
\eea
We prove the existence of the propagator for $Q_{bcd}$
by explicit calculating the inverse
operator in flat space-time limit.
In that limit one has
\bea
&& {\cal G}_{bcd}^{pqr}={\mathbf 1}_{bcd}^{pqr}\Box
              +{\cal K}_{bcd}^{pqr} \Box +{\cal B}_{bcd}^{pqr}\Box,\nn \\
&& {\cal K}_{bcd}^{pqr}=\dfrac{1}{\Box}\delta_b^{[q}\delta_{[c}^{p}\pro_{d]}\pro^{r]}, \nn \\
&& {\cal B}_{bcd}^{pqr}=\dfrac{1}{\Box}\delta_b^{p}\delta_{[c}^{[r}\pro_{d]}\pro^{q]},
\eea
where, $\Box\equiv\pro^2$, and we use the following rule for
anti-symmetrization over indices
$[a,b]\equiv \dfrac{1}{2}(ab-ba)$.
The propagator ${\cal G}^{-1}$ can be defined as a right or left inverse
operator to ${\cal G}$. We will choose a left operator,
i.e., ${\cal G}^{-1}{\cal G}={\mathbf 1}$.
The inverse operator reads
\bea
&&{\cal G}^{-1}=\dfrac{1}{\Box}({\mathbf 1}+4 {\cal K}-4 {\cal B} -6 {\cal F}), \nn \\
&&F_{bcd}^{pqr}=\dfrac{1}{\Box^2}\delta_{[c}^{[q}\pro_{d]}\pro^{r]}\pro_b\pro^p.
\eea

The inverse operator to the kinetic operator
${\cal G}^{(S)}_{ab}\equiv 3 \hD_a \hD_b+\eta_{ab} \hR$
for the field $S^a$ can be found for
the non-flat space-time with a constant Riemann  curvature
in a complete form
\bea
&& {({\cal G}^{(S)})}^{-1}_{ab}=\dfrac{1}{\hR}
\Big (\delta_{ab}-\hD_a \dfrac{1}{\hD \hD+\dfrac{\hR}{3}}\hD_b \Big ).
\eea
The inverse operator does not have additional poles,
but it has a singularity at $\hR \rightarrow 0$.
This is related to the fact of appearance of the additional
local symmetry (\ref{symtau}) in flat space-time limit.
Notice, that the curvature $\hR$ plays the role of
a mass scale which makes worse the ultra-violet
behavior of the propagator. Because of this
the model looks non-renormalizable
within the perturbative quantization scheme. One should notice that
the scale $\hR$ appears to be
a natural cut-off parameter related to the finite
size of the Universe. So that, the standard
perturbative technique of Feynman diagrams with unlimited internal momentum
inside loops needs some improvement at least.
Still it is possible that the model
can be renormalizable non-perturbatively,
since the original Lagrangian reveals high symmetry,
and, there is no dimensional coupling constants in the theory.

\section{Discussion}

In our previous paper \cite{pak} we have proposed a mechanism
of dynamical generation of Einstein gravity and cosmological
term via quantum corrections of torsion in the framework
of Yang-Mills type Lorentz gauge model.
The main ingredient of such a mechanism is the formation of
torsion vacuum condensate which we assumed to be
covariant constant
\bea
&&<\tilde R_{abcd}>= \difrac{1}{2} M^2(\eta_{ac} \eta_{bd}-\eta_{ad}\eta_{bc}),
\label{assumpn}
\eea
where $M^2$ is a mass scale parameter characterizing the
torsion condensate.
The factor $M^2$ need to be positive since it
corresponds to positive curvature space-time which can only be
created during the vacuum transition from the trivial vacuum
to the non-trivial one.
Expanding the original classical Lagrangian ${\cal L}$, (\ref{L0}),
around the new vacuum by shifting $\tilde R_{abcd} \rightarrow \tilde R_{abcd}
 + <\tilde R_{abcd}>$ one obtains the following torsionless part
of the Einstein-Hilbert effective action
\bea
&&{\cal L}_{EHeff}=-\difrac{1}{4} I_{GB}(\hR) - \difrac{1}{2} \hat R M^2
-\difrac{3}{2} M^4. \label{GBeff}
\eea
Notice, that the second and third terms in the effective
action are completely identical to those in the effective action
obtained from the Yang-Mills type Lagrangian \cite{pak}.
One should emphasize that even though the vacuum averaged value
$<\tilde R_{abcd}> $ is specified by (\ref{assumpn}), the classical
gravitational field $\hat R_{abcd}$ is not constrained in general.
The last term in the equation corresponds to
a positive vacuum energy density which is supposed to
be born during the vacuum transition.
This is consistent with the positiveness of the
scale parameter $M^2$ and with the fact
that the torsion condensate (\ref{assumpn})
corresponds to the gravitomagnetic part
of the Rieman-Cartan curvature \cite{pak}.

Notice that the torsion condensate (\ref{assumpn})
is supposed to be covariant constant. Rigorously speaking,
such a constant solution is unstable in general
like the constant chromomagnetic field configuration
in quantum chromodynamics which leads to vacuum
instability \cite{savv}. One can see that the ghost
kinetic operator in (\ref{tachyon}) is not positively
defined. This implies existence of tachyonic mode
in the theory, so that the constant curvature solution
corresponds to a false vacuum, and a true microscopic vacuum
should be realized by some other non-trivial solution.
In this respect the search of possible classical stable
vacuum solutions in Einstein-Gauss-Bonnet gravity
with torsion is of great importance \cite{canfora}.

In conclusion, we have considered a non-topological
Gauss-Bonnet type model of gravity with dynamical
torsion.
The model has a number of advantages to compare
with Yang-Mills type Lorentz gauge gravity.
In the absence of torsion the model reduces to a pure
topological Gauss-Bonnet gravity, i.e., one has a topological phase
where the metric is not specified a priori. The metric
obtains its dynamical content after
dynamical symmetry breaking in the phase of
effective Einstein gravity which is induced by quantum
torsion corrections.
Remarkably, the contortion in our model has only one physical spin two
field which can be interpreted as a torsion quantum counter-part to
the classical graviton. So that we don't have to quantize the metric
which can be treated as a classical object of the effective Einstein theory,
whereas the torsion (its spin two field component)
could be responsible for the quantum effects of gravitation.
Another interesting result is that the Hamiltonian part
corresponding to the spin two torsion component
is positively defined unlike the case of Yang-Mills type
Lorentz gauge theory. An obvious drawback of the model
is still the presence of the vector mode $K_d$ which produces a
negative contribution to the energy. One possible way
to remove this difficulty is to extend the model
by introducing supersymmetry
which supposed to be unbroken at scale near the Planckian one.
Possible applications of Lorentz gauge models
of gravity with Lagrangians quadratic in Riemann-Cartan curvature
in cosmology of early Universe will be considered elsewhere.

{\bf Acknowledgements}

One of authors (DGP) thanks Prof. M.L. Ge for useful discussions
and hospitality during visiting Chern Institute of Mathematics.

\end{document}